\documentclass[10pt]{iopart}

\usepackage{graphicx}
\usepackage{wrapfig}
\usepackage{epsfig}
\usepackage{subfigure}
\usepackage{dcolumn}% Align table columns on decimal point
\newcommand{\bm}{\bibitem}

\def\be {\begin{equation}}
\def\ee {\end{equation}}
\def\bea {\begin{eqnarray}}
\def\eea {\end{eqnarray}}
\def\nn {\nonumber}

\begin{document}
%\Large{Can collisional energy loss explain  
%nuclear suppression factor for light hadrons ? }
\title[]{Quenching of light hadrons at RHIC in a collisional energy 
loss scenario}
\author{Pradip  Roy$^a$\footnote{email: pradipk.roy@saha.ac.in}, 
Jan-e Alam$^b$, Abhee K. Dutt-Mazumder$^a$} 
\address{a) Saha Institute of Nuclear Physics, 1/AF Bidhannagar, Kolkata, 
India}
\address{b) Variable Energy Cyclotron Centre, 1/AF Bidhannagar, Kolkata, India}

\begin{abstract}
We evaluate the nuclear suppression factor, $R_{AA}(p_T)$ for light
hadrons by taking into account the collisional energy loss. We show that
in the measured $p_T$ domain of RHIC the elastic process is the dominant
mechanism for the partonic energy loss.

%{PACS numbers: 12.38.Mh, 24.85.+p, 25.75.-q,13.87.Fh }
%\\
%{Keywords: Collisional, energy loss, jets, Fokker-Planck}
\end{abstract}
%\maketitle
\section{Introduction}

%Recent measurements~\cite{npa757} of various probes at Relativistic
%Heavy Ion Collider (RHIC) provide direct
%indication of the presence of a novel phase possibly a Quark Gluon Plasma (QGP).
%Of these jet quenching has been thought to be the most promising one.
%The energy loss of the fragmenting partons causes depopulation of hadrons at 
%high transverse momentum
The  suppressions of high $p_T$ hadrons and  unbalanced 
back-to-back azimuthal correlations of the dijet events
in Au+Au collisions measured at  
RHIC~\cite{npa757} provide experimental evidence in
support of the jet quenching. Most of the calculations 
(see ~\cite{gyulassyreview} for a review) 
considers the energy loss due to induced bremsstrahlung radiation and 
reproduces the observed nuclear suppression
of light hadrons ($\pi, \eta$) in $Au+Au$ collisions for centre of mass
energy $\sqrt{s_{\rm{NN}}}=62-200$ GeV at RHIC.
The effect of collisional loss  was completely ignored in most of
the previous calculations. 
However, the non-photonic single electron spectrum from heavy meson decays 
measured by
PHENIX Collaboration~\cite{raacharm} shows much larger suppression
than expected. By considering radiative energy loss for heavy quarks the
data cannot be reproduced as radiation is suppressed for heavy quarks
due to dead-cone effects. Thus, there has been a renewed interest to
revisit the importance of collisional energy loss both for light as well
as heavy quarks. 
%\cite{speigne} 
%finite size medium the collisional energy loss is suppressed. However, they
%could not seperate out the individual contributions. This finding
%has been contrasted by Djordjevic~\cite{djordjevic06} where it is shown that
%finite-size effects are negligible. Adil et al.~\cite{PRC752007} has
%found similar results. Peshier~\cite{PRC} has recently shown that at 
%RHIC energies collisional energy loss becomes probable which is similar
%to the results obtained in Ref.~\cite{abhee05}. 

The partonic energy loss due to collisional
processes was revisited in~\cite{abhee05} and its importance
was demonstrated in the  context of RHIC  
in\cite{roy06}. 
It is the purpose of the present work to show that the omission of the
collisional energy loss to explain the RHIC data is not justified.
To this end, we calculate the nuclear suppression factor $(R_{AA})$
for pions ($\eta$) considering only the collision energy loss. 

\section{Theoretical framework}
In order to calculate the $p_T$ distribution of hadrons from parton fragmentation
we need the phase space distribution of partons, $f(\bf p,t)$. The dynamical evolution of 
$f(\bf p,t)$  
is obtained by solving the 
Fokker-Planck (FP) equation which reads,
\bea
\left (\frac{\partial}{\partial t}
-\frac{p_\parallel}{t}\frac{\partial}{\partial p_\parallel}\right )f({\bf p},t)
&&=\frac{\partial}{\partial p_i}[p_i\eta f({\bf p},t)]\nn\\ 
&&+\frac{1}{2}
\frac{\partial^2}{\partial p_\parallel^2 }[{B_\parallel}({\bf p})f({\bf p},t)]
\nn\\
&&+\frac{1}{2}
\frac{\partial^2}{\partial p_\perp^2}[{B_\perp}f({\bf p},t)],
\label{fpexp}
\eea
where the second term on the left hand side arises due to 
expansion~\cite{baym}. 
Bjorken hydrodynamical model~\cite{bjorken1983}  
has been used here for space time evolution.
In Eq.~(\ref{fpexp})  $f({\bf p},t)$ represents the  
distribution function of the partons under study, 
$\eta=(1/E)dE/dx$,
denotes drag coefficient,
$B_\parallel=d\langle(\Delta p_\parallel)^2\rangle/dt$,
$B_\perp=d\langle(\Delta p_\perp)^2\rangle/dt$, 
represent diffusion constants along parallel and
perpendicular directions of the propagating partons.

The  matrix elements required to calculate the transports coefficients include
diagrams involving exchange of massless gluons which render
$dE/dx$ and $B_{\parallel,\perp}$ infrared divergent. Such
divergences can naturally be cured by using the hard thermal
loop (HTL)~\cite{pisarski} corrected propagator for the gluons, i.e.
the divergence is shielded by plasma effects. 
For jet with energy $E >> T$ (see ~\cite{abhee05} for details) the energy 
loss is given by
\begin{equation}
\frac{dE}{dx} \sim \alpha_s^2\,T^2\,C_R \ln\frac{E}{g^2T}
\end{equation}

Having known the drag and diffusion~\cite{abhee05}, we solve the FP equation
 using Green's function techniques with the initial condition
\begin{equation}
P(\vec{p},t=t_i|\vec{p_0},t_i) = \delta^{(3)}(\vec{p}-\vec{p_0})
\end{equation}
along the line of Refs.~\cite{moore05,rapp}.

The solution with an arbitrary initial momentum distribution
can now be written as~\cite{moore05,rapp}, 
\begin{equation}
E\frac{dN}{d^3p^j}|_{y=0} = \int\,d^3p_0^j\,
P(p^j,t|p_0^j,t_i)
E_0\frac{dN}{d^3p_0^j}|_{y_0=0}\,
\end{equation}
where $j$ stands for any parton species. 

\begin{figure}
\centerline{\epsfig{figure=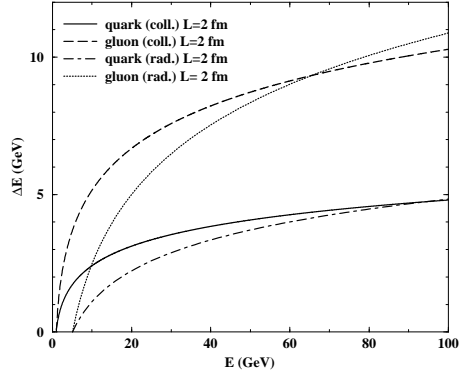, height= 5 cm}}
\caption
{Collisional versus radiative energy loss. The radiative loss is taken from
\protect\cite{gyulassy00prl} (see~\protect\cite{abhee05} for details)}
\label{fig1b}
\end{figure}

In order to take into account the jet production geometry
we assume that all the jets are not produced at the same
point and the path length traversed by these partons before
fragmentation are not the same. 
It is also assumed that
the jet initially produced at $(r,\phi)$
leave the plasma after a proper time ($t_L$) or equivalently after 
traversing a distance $L$ (for light quarks $t_L \sim L$) given by 
$t_L=\sqrt{R^2-r^2\sin{\phi}^2}-R\cos{\phi}$. 
As this is not a measurable quantity, we have to average it out to
obtain the $p_T$ spectra of hadrons: 
\bea
\frac{dN^{\pi^0(\eta)}}{d^2p_Tdy} &=& \sum_f\,\int\,d^2r{\cal P}(r)\,
\int_{t_i}^{t_L}\,\frac{dt}{t_L-t_i}\int\,
\frac{dz}{z^2}\nonumber\\
&&\times\,D_{\pi^0(\eta)/j}(z,Q^2)|_{z=p_T/p_T^j}
\,E\frac{dN}{d^3p^f}|_{y_j=0},
\eea
%Solving the FP equation with the boundary conditions, 
%$f(\vec{p},t)\rightarrow 0$ for $|p|\rightarrow \infty$,
%we evaluate 
where $t_c$ is the time when temperature cools down to the transition 
temperature  $T_c$ (=190 MeV)~\cite{katz}. 
The temperature profile is taken as in Ref.~\cite{moore05}.
The nuclear suppression factor, $R_{AA}$ is defined as 
\bea
R_{AA}(p_T) 
= \frac{\frac{dN_{AA}^{\pi^0(\eta)}}{d^2p_Tdy}}
{\left[\frac{dN_{AA}^{\pi^0(\eta)}}{d^2p_Tdy}\right]_0}
%%= \frac{\sum_a \int f_a({\bf p^{\prime}}, \tau_c)|_{p_T^{\prime} = p_T/z}
%%D_{a/\pi^0}(z,Q^2)dz}{\sum_a \int f_a({\bf p^{\prime}},\tau_i)
%%|_{p_T^{\prime} = p_T/z}, 
%%D_{a/\pi^0}(z,Q^2)dz}
%%\tau_i)D_{\pi^0/i}(z,Q^2)dz}
\eea
where the suffix `0' in the denominator indicates that energy loss
has not been considered while evaluating the expression.

\section{Results}

To understand the relative importance of the energy loss mechanisms we
plot the contributions from radiative and collisional processes
in Fig.~\ref{fig1b}.
It is observed that collisional energy loss is the dominant 
mechanism of energy loss for 
parton energy up to $E=E_c \sim 85 (60) $ GeV for quark (gluon).

%\vskip .5in

\begin{figure}
\centerline{\epsfig{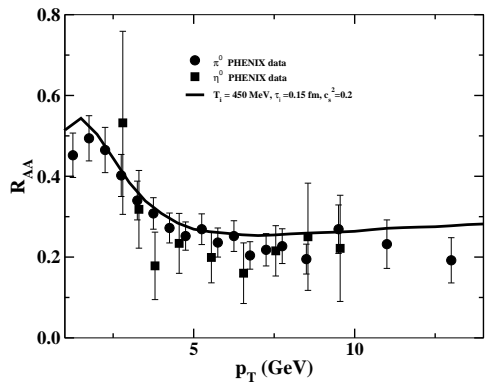}}
\caption
{Nuclear suppression factor for pion. Experimental data are taken from
PHENIX collaboration \cite{nuclex06} 
for Au + Au collisions
at $\sqrt{s}=200$ GeV. Solid  line indicates result from the
present calculation with collisional energy loss of the partons
propagating through the plasma 
before fragmenting into pions.
We have taken $T_i$=450 MeV, $t_i$=0.15 fm and $c_s^2$=0.2
}
\label{fig1}
\end{figure}

Nuclear suppression factor, $R_{AA}$ for neutral pions is plotted as
a function of transverse momentum in 
Fig.~\ref{fig1} which describes the PHENIX data\cite{nuclex06} for $Au+Au$ at
$\sqrt{s}=200$ GeV reasonably well. 
It should be noted  here that the $R_{AA}(p_T)$
with collisional loss has a tendency to increase
for higher $p_T$, indicating diminishing importance of 
collisional loss at this domain, where the radiative
loss may become important. Therefore, a detailed
calculation with both collisional and radiative 
loss may be useful to delineate the importance
of individual mechanism.

\begin{figure}
\centerline{\epsfig{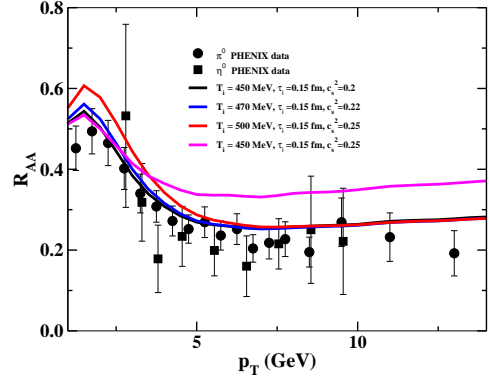}}
\caption
{Same as Fig.(\ref{fig1}) for various values of initial
temperature, sound velocity and thermalization time
}
\label{fig1a}
\end{figure}

It is important to note that the result for $R_{\rm{AA}}$ is very sensitive to 
the initial temperature ($T_i$), equation of state(EOS) (through velocity of sound ($c_s$)) and the 
thermalization time ($t_i$). This is demonstrated in 
Fig.~(\ref{fig1a}). This aspect, has not been received much attention
in the literature. Our calculations show that the data can be reproduced
reasonably well with $T_i$ = 450 MeV, $c_s^2$=0.2 
and $t_i$=0.15 fm/c similar to those required to reproduce the single
photon data~\cite{JJPAB}.

\section{Summary}

In conclusion,
we have used the $R_{AA}(p_T)$ for $\pi^0$ measured
by PHENIX collaboration to characterize the QCD medium
created after the Au + Au collisions at RHIC. The data
can be be explained with $T_i\sim 450 - 500$ MeV depending
upon the EOS, {\it i.e.} velocity of sound, $c_s$ in the medium.
For a lower value of $c_s$ the expansion is slow. Consequently, the
energy loss process occurs for a longer duration in the medium
for a given $T_c$. Therefore, for smaller values of $c_s$ the
corresponding $T_i$ is smaller. 
%For SPS energies
%considered here the value of $T_i$ obtained from the
%present analysis is close to the value estimated 
%from the analysis of low $p_T$ direct photon 
%spectra~\cite{spsphoton1,spsphoton2}.

Our investigations suggest that in the measured
$p_T$ range of light hadrons at RHIC
collisional, rather than the radiative, is the dominant mechanism of 
jet quenching. This is in sharp contrast to all the previous analyses.

\end{document}